\documentclass[aps,prl,twocolumn,floatfix,tightenlines,showpacs,notitlepage]{revtex4-1}
\usepackage{amsfonts} 
\usepackage{amsmath} 
\usepackage{amssymb}
\usepackage{blindtext}
\usepackage{lineno}
\usepackage{cancel}
\usepackage{bm}
\usepackage[usenames, dvipsnames]{color} 
\usepackage{dcolumn}
\usepackage{epic} 
\usepackage{epsfig}
\usepackage{eucal} 
\usepackage{esdiff}
\usepackage{graphicx}
\usepackage[dvipsnames]{xcolor}
\definecolor{darkgreen}{HTML}{005A00} 

\usepackage[breaklinks,colorlinks = true,linkcolor = blue,urlcolor=blue,citecolor=blue]{hyperref}

\begin{document}
\title{Turbulent flows are not \textit{uniformly} multifractal}
\author{Siddhartha Mukherjee} 
\email{siddhartha.m@icts.res.in}
\author{Sugan D. Murugan} 
\email{sugan.murugan@icts.res.in}
\author{Ritwik Mukherjee} 
\email{ritwik.mukherjee@icts.res.in}
\author{Samriddhi Sankar Ray} 
\email{samriddhisankarray@gmail.com}
\affiliation{International Centre for Theoretical Sciences, Tata Institute of Fundamental Research, Bengaluru 560089, India}
\begin{abstract} 


	The Frisch-Parisi multifractal formalism remains the most compelling
	rationalisation for anomalous scaling in fully developed turbulence. We
	now show that this formalism can be adapted \textit{locally} to
reveal the spatial distribution of generalized dimensions and of how multifractal the energy
	dissipation field is. In particular, we show that most regions of the
	flow are close to being mono-fractal and these are interspersed with
	islands of multifractality corresponding to the most singular
	structures in the flow. By defining a suitable measure $\Phi ({\bf x})$
	of the spatial variation of multifractality, we show that this grows
	logarithmically with the extent to which the energy dissipation varies
	locally around ${\bf x}$. These results suggest ways to understand how
	singularities could arise in disparate regions of a flow and provides
	new directions in understanding anomalous dissipation and
	intermittency. We then employ the same technique to a non-intermittent,
	model turbulent flow to check the robustness of our conclusions.

\end{abstract}
\date{13\textsuperscript{th} July, 2023}
\maketitle

A central scaffold for interpreting and describing out-of-equilibrium pattern
forming
processes~\citep{stanley1988multifractal,procaccia1988universal,cross1993pattern,shlesinger1993strange},
multifractality has inevitably woven itself into turbulence
theory~\citep{sreenivasan1991fractals,boffetta2008twenty,dubrulle2022correspondence},
phenomenology~\citep{argoul1989wavelet} and data
analysis~\cite{muzy1991wavelets}.  The Frisch-Parisi multifractal
model~\cite{Frisch-Parisi,Frisch-Book} indeed remains the most powerful
theoretical justification for the statistical properties of turbulence like the
anomalous scaling of correlation functions of velocity
differences~\cite{Frisch-Book,Mitra2004,Mitra2005,Ray-NJP,Ray-PRL,RathorPRE}, strongly
non-Gaussian distributions of velocity
gradients~\cite{Nelkin,KailasnathPRL,benzi2009fully} and fluid
accelerations~\cite{Porta-Nature}.  Constructed on the premise of an
intermittent, infinite Reynolds number flow with a form of \textit{local}
scale-invariance under the transformations $r \to \lambda r$, $u\to \lambda^h
u$ and $t\to \lambda^{1-h}t$~\citep{frisch1991global}, it provides corrections
to the Kolmogorov \textit{mean field} theory in a way which is consistent with
measurements made in experiments and direct numerical simulations (DNSs).  The
key ingredient in the construction of such theories is the assumption that
intermittency effects lead to a range of H\"older exponents $h \in [h_{\rm
min}, h_{\rm max}]$ for the (turbulent) velocity field ${\bf u}({\bf x})$. The
simplest interpretation for these exponents relate to the (inertial range)
scale-invariance in turbulence of the longitudinal velocity difference $\delta
u_r \equiv \langle {\bf u}({\bf x} + {\bf r}) - {\bf u}({\bf x}) \rangle \sim
r^h$~\cite{Frisch-Book}.  

These ideas, when applied to the intermittent energy
dissipation field $\epsilon ({\bf x})$, leads to the remarkable result that the total dissipation in $d$-dimensional ``boxes'' of size $r$, denoted $\mathcal{E}_r$, scales as a
\textit{fractal} power-law with a variable scaling exponent $\alpha$ as $\mathcal{E}_r \sim r^{\alpha - 1 + d}$~\cite{Frisch-Book,MS-Nucl,MS-PRL,MS-JFM}. This is
a direct consequence of the multifractal interpretation that despite the
three-dimensional embedding dimension, the energy dissipation --- which is a
culmination of the energy cascading process --- accumulates in different,
entangled fractal subsets with unique dimensions. It is then possible to
associate the fractal dimension $f_\alpha$ of these subsets with exponents
lying between $\alpha$ and $\alpha + d\alpha$ yielding the well-known singularity or
multifractal spectrum $f_\alpha - \alpha$. 

The framework of these ideas are a powerful tool bridging the conceptual
picture of the energy cascade with intermittency of the dissipation field.
Indeed, it is easy to show that $\alpha = 3h$ is exactly the same (arbitrary)
scaling exponent which leaves the Navier-Stokes equations invariant under its rescaling transformations~\citep{Frisch-Book}.  Furthermore, within the Kolmogorov non-intermittent
phenomenology, there is a single exponent $h = 1/3$ (i.e. $\alpha = 1$)~\citep{frisch1991global,benzi1984multifractal} which leads to the familiar 2/3
and 5/3 laws of turbulence. However, obtaining the measured exponents and the corrections to the  Kolmogorov prediction from the Navier-Stokes equation 
still remains elusive.

This singularity spectrum though remains a central pillar in modern statistical theories 
of fully developed turbulence. Beginning with the earliest measurements of Meneveau and 
Sreenivasan~\cite{MS-Nucl, MS-PRL, MS-JFM}, the robustness of the multifractal 
nature of the kinetic energy dissipation field has never been in question. And yet, 
all these measurements pertain to the statistics of the \textit{entire} field (or signal). This is somewhat 
surprising because implicit in the ideas of multifractality is the spatial \textit{fluctuation} of the 
scaling exponents over the flow field. Is it possible 
then to actually probe the multifractal nature of turbulence in a local  way, i.e., 
to have estimates of the spatial dependence of the generalized dimensions $D_q({\bf x})$,
the singularity spectrum $f_\alpha({\bf x})-\alpha({\bf x})$ and thence of course the 
distribution of the H\"older exponents $h({\bf x})$?  
The closest, so far, have been recent attempts~\cite{dubrulle2017,dubrulle2019,dubrulle2019JFM,dubrulle2020} using wavelet techniques and local energy transfer concepts to characterize fields similar to the intractable $h({\bf x})$. While characterizing H\"older exponents point-wise may indeed be difficult, a local multifractal analysis, keeping the robustness of the Frisch-Parisi formalism, opens up a way to reveal the crucial underlying variation in multifractality, which has so far remained uncharted.

\begin{figure}
	\includegraphics[width=1.0\columnwidth]{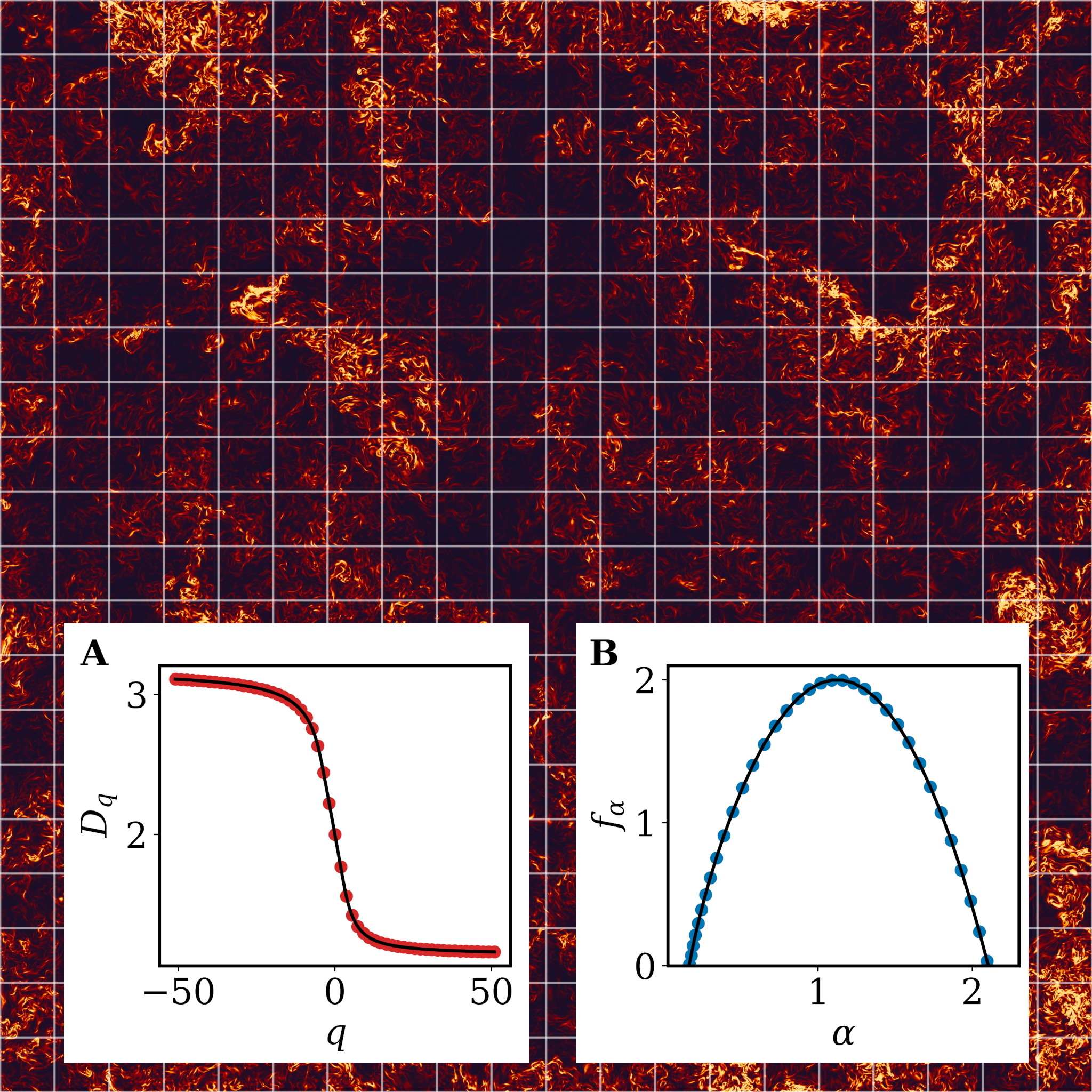}
	\caption{\textbf{Multifractal Dissipation Field.} Pseudo-color plot of dissipation $\epsilon$ at a representative cross-section from the Johns Hopkins isotropic $4096^3$ data. Superimposed on this is a regular tiling as a pictorial guide to show how the local multifractal properties are calculated (made with Processing~\citep{reas2007processing,pearson2011generative}). 
	(Inset) \textbf{(A)} Generalized dimensions $D_q$ and \textbf{(B)} the singularity spectrum $f_\alpha - \alpha$ calculated over the full 2D cross-section shows the essential multifractal nature of the dissipation field, consistent with earlier measurements.}
	\label{fig:eps}
    \end{figure}

In this paper we show how a locally adapted
construction of multifractal measures of turbulence dissipation fields throws
up surprises. In particular, we find that \textit{most} of the flow is essentially monofractal with an almost delta
function for $f_\alpha$ at $\alpha \approx 1$ which corresponds to the Kolmogorov
mean field exponent $h = 1/3$. The few patches of multifractal behaviour (with
broad $f_\alpha({\bf x})-\alpha({\bf x})$ curves) are directly correlated with
spatial regions of enhanced dissipation and, by extension, intermittency.
Indeed, a more accurate description of fully developed turbulence would be intermittent, multifractal islands on a vast and calm Kolmogorovean sea.

Our analysis is based on kinetic energy dissipation fields $\epsilon ({\bf x})
\equiv 2\nu S_{ij}S_{ij}$, where $S_{ij}$ is the symmetric part of the velocity gradient tensor,
obtained from 3 different direct numerical simulations, with very different
(Taylor-scale based) Reynolds numbers $Re_\lambda$, of the three-dimensional
(3D), triply-periodic, incompressible Navier-Stokes equation.  For the smallest
$Re_\lambda \approx 200$, we use our own fully de-aliased pseudospectral code
with $N = 512^3$ collocation points  and a constant energy-injection rate on
the first 2 shells. For higher Reynolds numbers, we use publicly available data
from the Johns Hopkins Turbulence Database (JHTD)~\cite{perlman2007data,li2008public,yeung2012dissipation} with $N = 1024^3$
($Re_\lambda \approx 433$) and $4096^3$ ($Re_\lambda \approx 610$). Our results
are consistent across this wide range of Reynolds numbers and independent of
simulations; in what follows, we present converged results, using a $4096^2 \times 192$ subset of the $4096^3$ dataset.

Let us first recall how the classical multifractal spectrum for a $d$-dimensional
dissipation field is constructed. Denoting the average dissipation on a scale
$r$ by $\epsilon_r$, we construct $N_r$ number of (space-filling)
$d$-dimensional boxes of size $r$ in the full domain. This allows us to
estimate the total dissipation $\mathcal{E}_r \sim \epsilon_r r^d \sim
r^{\alpha - 1 + d}$ within each box. By taking
the $q-$th moment of this and summing over all $N_r$ boxes one obtains the
\textit{partition function} $Z_q \equiv \sum_{N_r} \mathcal{E}_r^q \sim
r^{(q-1)D_q}$, where $D_q$ is the generalized dimension and the following
relation holds: $\sum_{N_r} r^{(\alpha - 1 + d)q} \sim r^{(q-1)D_q}$. Further analysis~\cite{Frisch-Book,MS-Nucl} yields the exact relations for the singularity spectrum:
\begin{align}
	\alpha &= \frac{d}{dq}[(q-1)(D_q -d + 1)]\\
	f_\alpha &= \alpha q - (q-1)(D_q - d + 1) + d-1
\end{align}
Clearly, within the mean field, monofractal Kolmogorov ideas for 3D turbulence, $D_q = d = 3$ leading to 
$\alpha = 1$ ($h = 1/3$) and $f_\alpha = d = 3$ and thence a $\delta$-function like $f_\alpha - \alpha$ curve. 

In Fig.~\ref{fig:eps} we show a representative plot of a 2D ($d = 2$) slice of
the $\epsilon$ field constructed from the $4096^3$ JHTD data; the insets
show the generalized dimensions $D_q$ and the singularity spectrum $f_\alpha - \alpha$ for this slice of data,
which is consistent with results reported earlier~\cite{Frisch-Book,MS-Nucl}: The broad $f_\alpha$ curve 
is the most precise indicator that turbulence admits a range of scaling exponents and not just the mean field Kolmogorov exponent $\alpha = 1$.

However, as is well known --- and illustrated in Fig.~\ref{fig:eps} --- the
dissipation field is strongly intermittent. It is not immediately obvious, therefore, whether the fractal sets on which $\epsilon({\bf x})$ is distributed are themselves uniform in space. There might, in fact, be an equally strong variation in multifractality over ${\bf x}$, which would be revealed if it were possible to measure the generalized dimensions $D_q({\bf x})$ and $f_\alpha ({\bf x}) - \alpha ({\bf x})$ locally.
These variations would not only help 
connect the ideas of the cascade with multifractality but also provide important insights 
in the detection of (possible) singular $h < 1/3$ regions with anomalous dissipation~\citep{onsager1949statistical,eyink2006onsager}.
Important and enticing as this question is, the very nature of the multifractal calculation precludes any 
possibility of a \textit{single} point ${\bf x}$ measurement of such quantities. This is because at a practical level  
the partition function $Z_q({\bf x})$ must be measured over a range of scales to extract the generalized 
dimensions $D_q({\bf x})$ from which follows the (local) singularity spectrum and scaling exponents.

\begin{figure}
	\includegraphics[width=1.0\columnwidth]{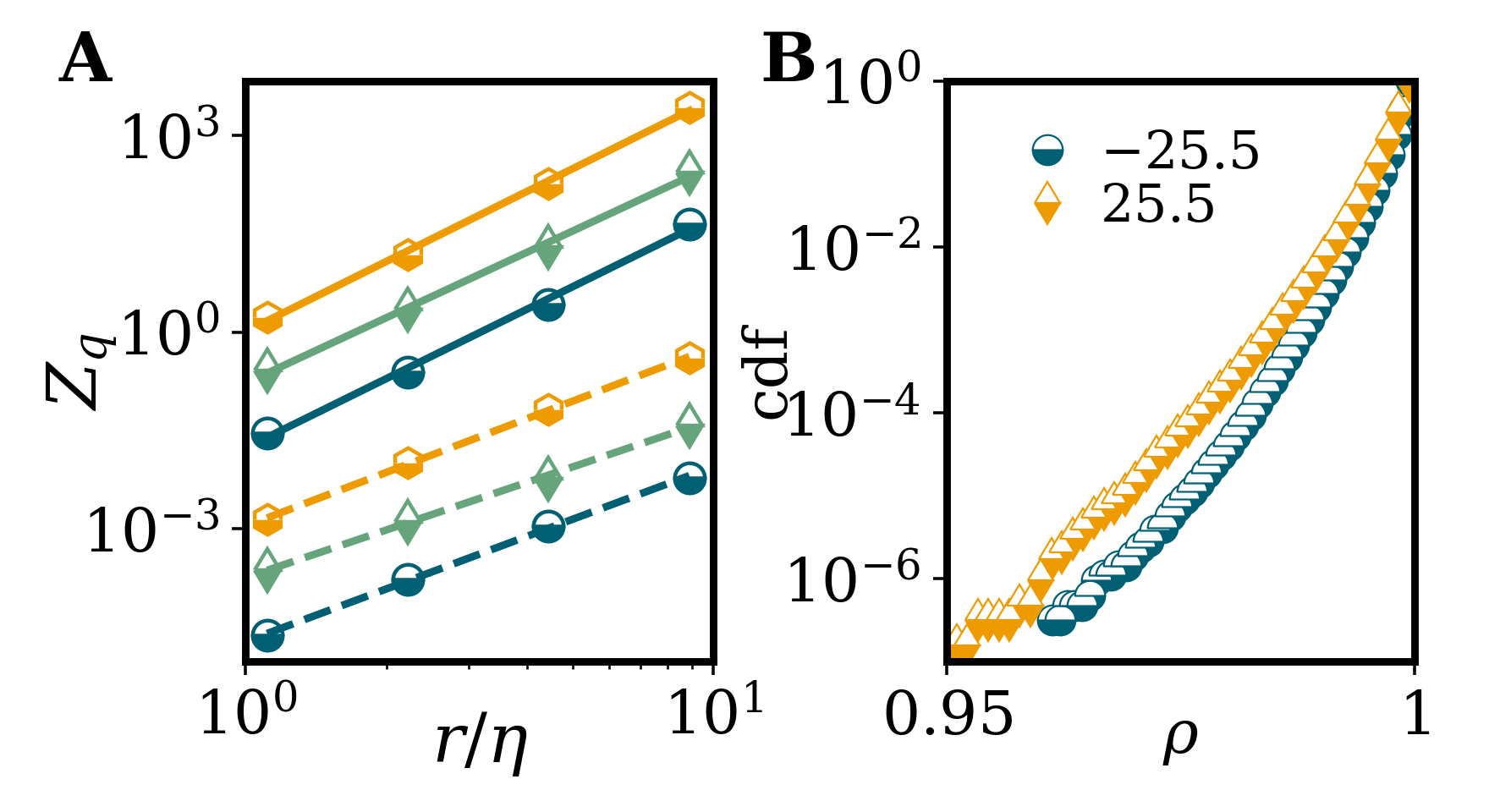}
	\caption{\textbf{Local Partition Function.} \textbf{(A)} Loglog plots of the partition function $Z_q({\bf x})$ vs $r$ for $q = -25.5$ (solid lines) and $q= 25.5$ (dashed lines), vertically shifted for clarity, calculated in randomly chosen spatial locations in data-tiles of size $\approx 10\eta$. \textbf{(B)} Cumulative distribution function of the Peason correlation coefficient $\rho$ for the linear-regression fits used to obtain $D_q$, for $q=-25.5$ and $q=25.5$, over all tiles.}
	\label{fig:localZq}
\end{figure} 

To circumvent this problem, we develop a tiling approach to allow us
to self-consistently measure the spatial variation in the multifractality of
the field. This is illustrated in Fig.~\ref{fig:eps} where a (exaggerated)
white grid is superimposed on the 2D slice of the dissipation field, leading to square data-tiles with egde $\mathcal{L}_{T}$ (or cubical divisions used for 3D analysis). We then
treat each of these tiles, centered at ${\bf x}$, independently and calculate the multifractal measures
in them as one would ordinarily for the full domain. The size of these tiles
was tested in the range between $2\eta < \mathcal{L}_T < 16\eta$, where $\eta$ is the Kolmogorov
dissipation scale. Larger tiles, which are still smaller than the inertial range, give a wider range of $r$ values over which to construct $Z_q({\bf x})$ while, for an ideal local measure, 
we would like the tiles to be as small as possible. However, the lower end of $\mathcal{L}_T$ is dictated by the constraint that 
we need \textit{enough} points to measure the scaling of $Z_q({\bf x})$ unambiguously. At an
operational level, this is not obvious since, unlike the full domain, we have fewer points within individual tiles on which the measurement can be made. 

Calculating the local variation of $D_q ({\bf x})$ with $q$, requires obtaining a clean scaling of the partition function $Z_q ({\bf x})$ now measured within the tiles. Thus, in Fig.~\ref{fig:localZq}(A), we begin by showing representative plots of the partition function for $q = -25.5$ (solid lines) and $q = 25.5$ (dashed lines), calculated in three-dimensional tiles with $\mathcal{L}_T \approx 10\eta$, at randomly chosen ${\bf x}$ locations. While this plot already suggests that a $Z_q$ vs $r$ scaling can be realiably obtained, we test the overall accuracy of doing so in Fig.~\ref{fig:localZq}(B) by calculating the cumulative distribution of the Pearson correlation coefficient $\rho$ for linear-regression fits used to obtain $D_q$, over all tiles. The distribution shows a high degree of confidence, with more than $99.98\%$ of tiles with $\rho>0.98$, for both $q$ values.

\begin{figure}
	\includegraphics[width=1.0\columnwidth]{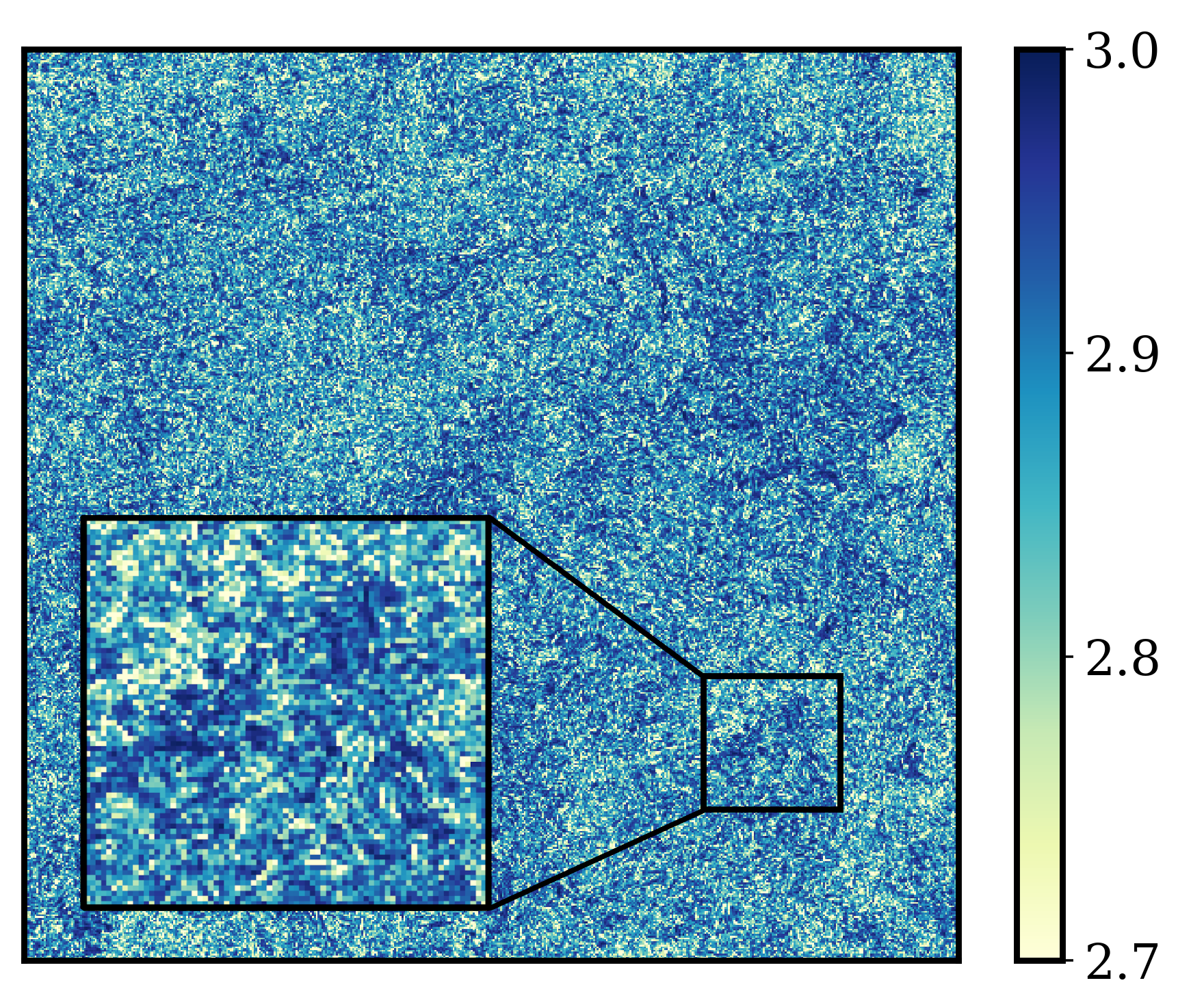}
	\caption{\textbf{Correlation Dimension Field.} Our local multifractal analysis makes it possible to calculate fields of generalized dimensions. We show a cross-section of the $D_2$ field, also known as the correlation dimension. Our analysis coarsens a $4096\times 4096\times 8$ data-slice to $512 \times 512$ tiles. This reveals a stark variation in $D_2$ over space, which remained hitherto unseen. Coherent patches of similarly valued regions of $D_2$ are found nestling in a fluctuating (and non-random) field.}
	\label{fig:D2}
\end{figure}

Clearly, these plots show that a local $D_q({\bf x})$ can be meaningfully extracted by using the prescription we propose. We found our results to be insensitive to the
chosen $\mathcal{L}_T$, indicating the consistency and convergence of our approach. In what follows, we report results from a tiling of $\mathcal{L}_T \approx 10\eta$ and carry out the analysis in three dimensions. Before stepping into spatially varying multifractal spectra, we pause to look at the special case of $D_q$ for $q=2$, also known as the correlation dimension, that provides a measure of inhomogeneity in a fractal set~\cite{grassberger1983characterization,paladin1987anomalous}, or simply $D_2$. Fig.~\ref{fig:D2} shows a planar cross-section of the $D_2({\bf x})$ field, starkly varying in space, with sizeable pockets of coherent regions of similarly valued correlation dimensions. This also shows that the field is far from random, and at some level this is reflective of the structures in the dissipation field. We wish to underline that our method allows, perhaps for the first time, to visualize this field, which further opens up directions to study the structure of these intrinsic, and as yet elusive, features of turbulence.

\begin{figure*}
	\includegraphics[width=0.328\linewidth]{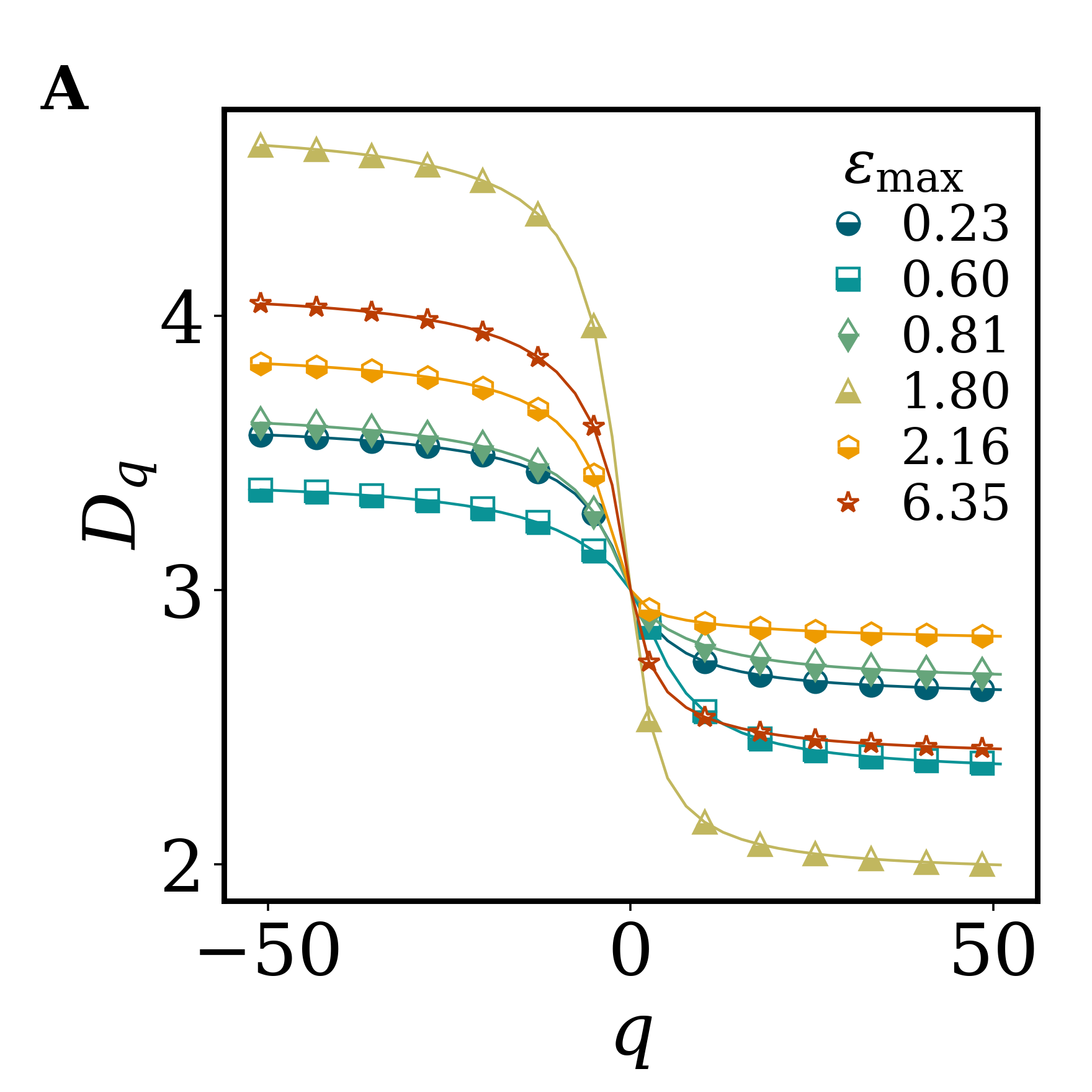}
	\includegraphics[width=0.328\linewidth]{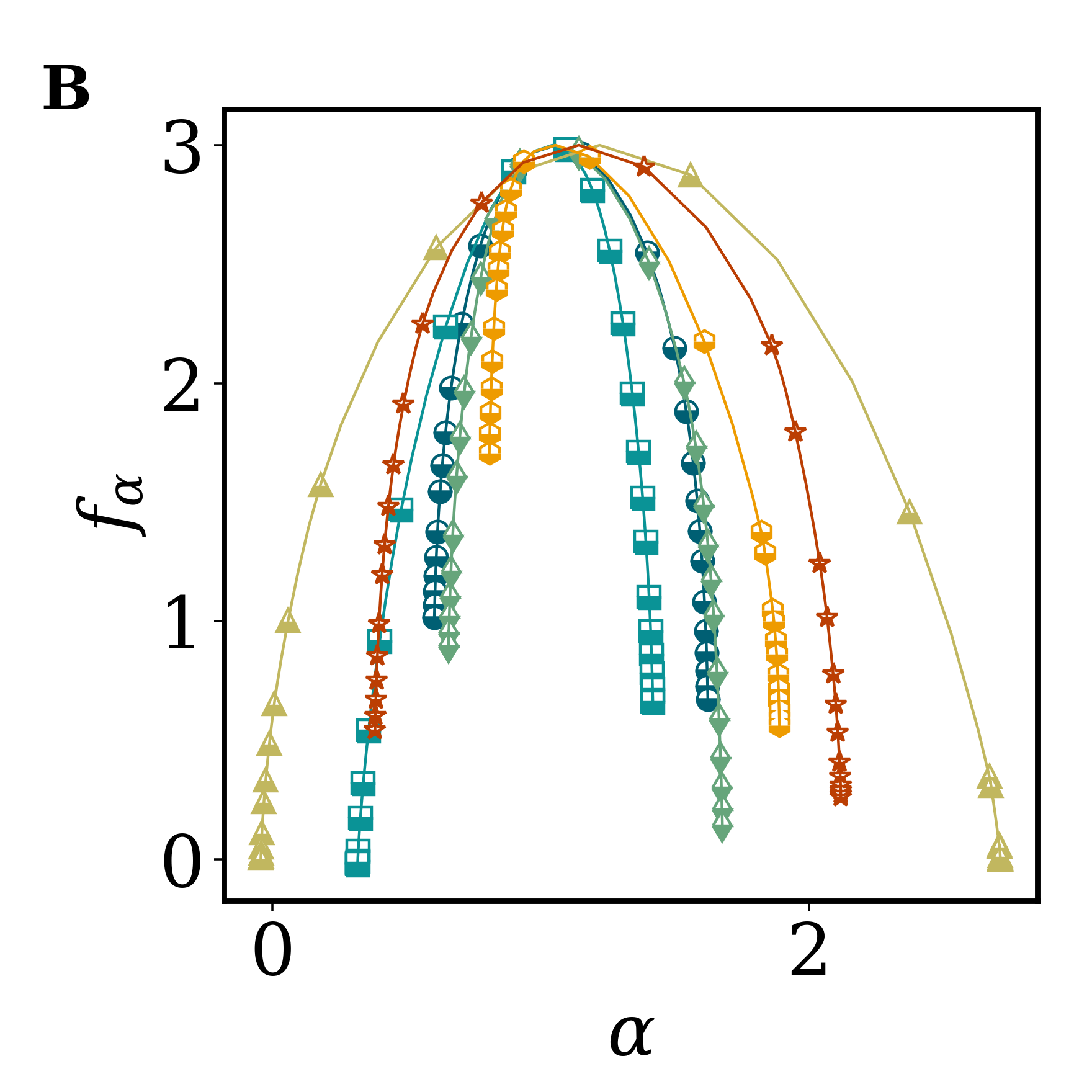}
	\includegraphics[width=0.328\linewidth]{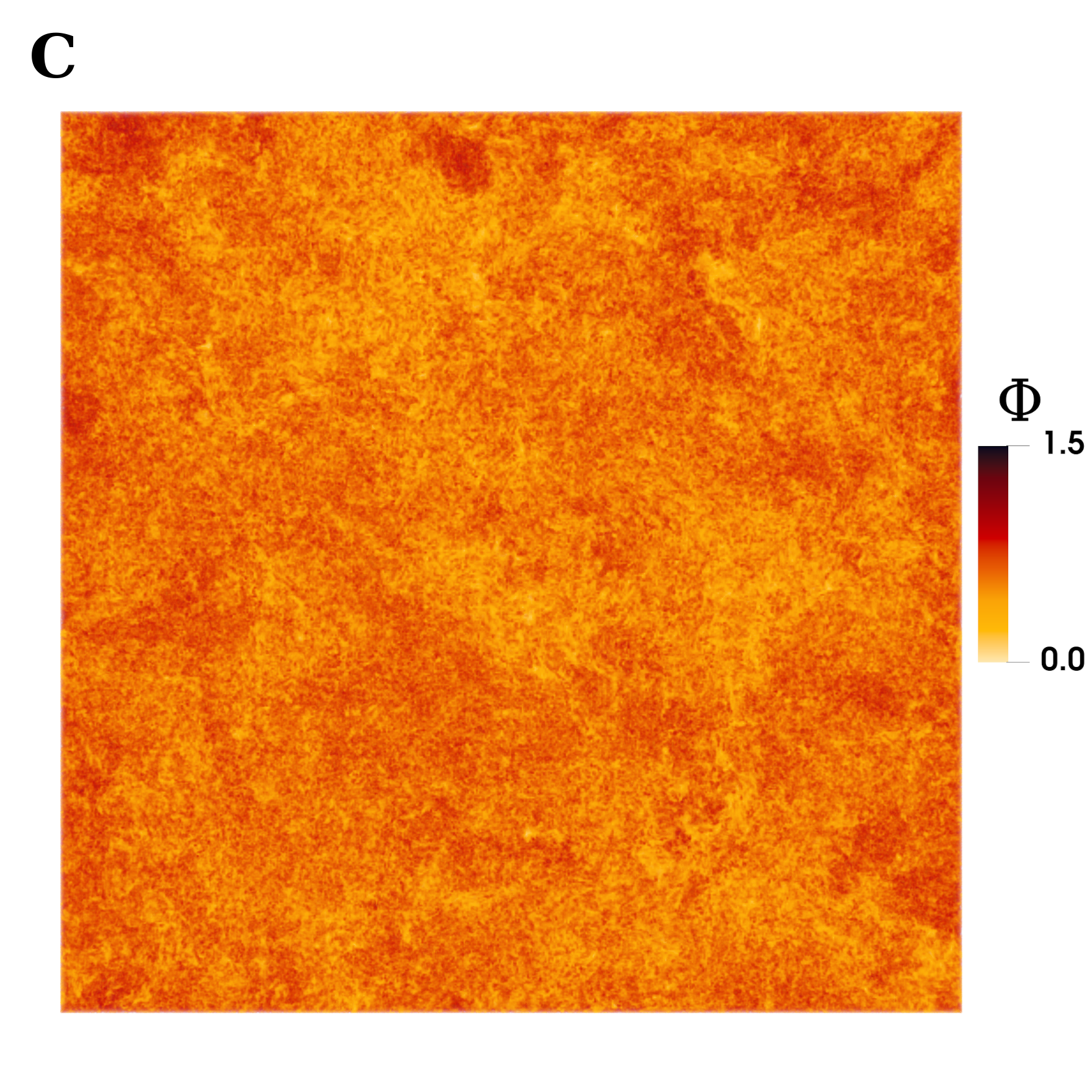}	
	\includegraphics[width=0.328\linewidth]{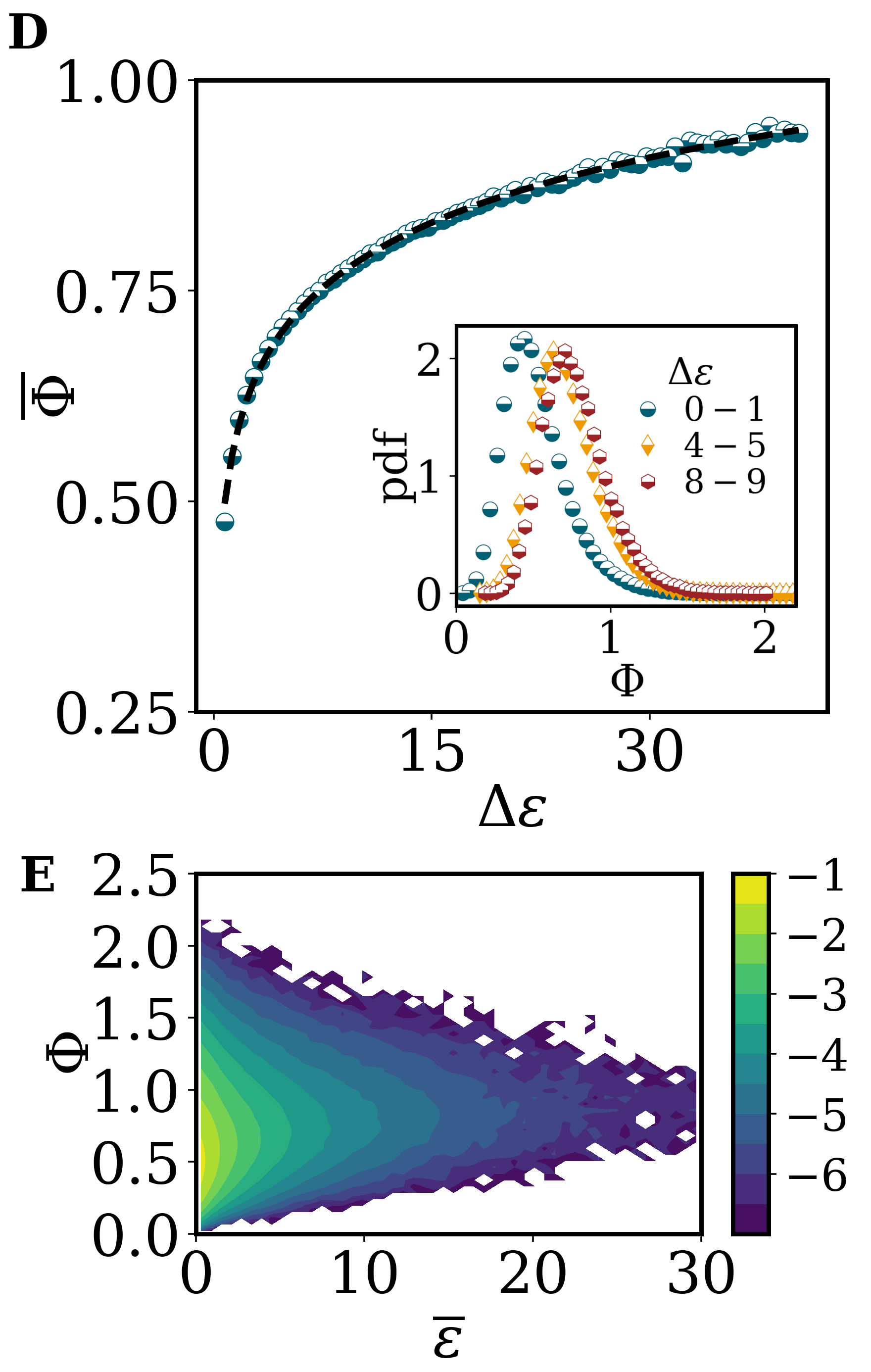}
	\includegraphics[width=0.65\linewidth]{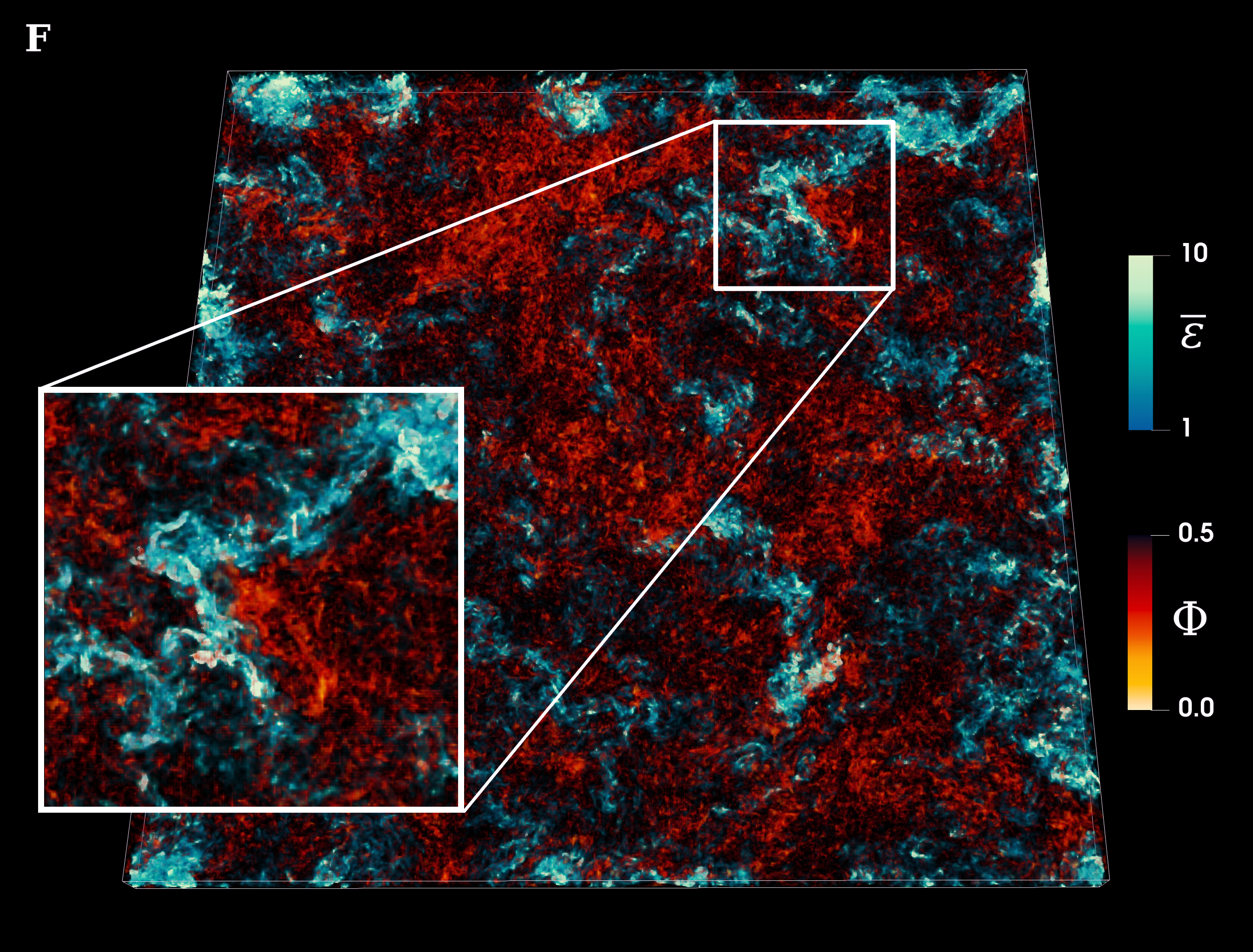}
		\caption{\textbf{Local Multifractality.} \textbf{(A)} Generalized dimensions $D_q$ vs $q$, calculated locally for randomly sampled tiles, together with \textbf{(B)} the corresponding $f_\alpha-\alpha$ spectra of singularity strengths, show a stark variation in the multifractal properties between different spatial regions. \textbf{(C)} The $\Phi$ field gives a strong visual impression of this variation, showing that large regions of the flow are actually almost monofractal with $\Phi \approx 0$, with pockets of $\Phi \sim \mathcal{O}(1)$. \textbf{(D)} (inset) shows the pdf of $\Phi$ shifting towards higher $\Phi$ when sampled in regions of higher $\Delta \epsilon = \epsilon_{\rm max} - \epsilon_{\min}$ (legend in multiples of the global mean $\langle \epsilon \rangle$). \textbf{(D)} shows the mean value of $\Phi$ in fact grows as the logarithm of $\Delta \epsilon$. \textbf{(E)} shows a full joint-distribution of $\Phi$ and the mean dissipation $\overline{\epsilon}$ in each tile. While lower $\overline{\epsilon}$ corresponds most likely to low values of $\Phi$, an increase in $\overline{\epsilon}$ makes larger $\Phi$ equally likely, while also increasing the smallest admissible $\Phi$ value. \textbf{(F)} A volume-rendering of $\overline{\epsilon} \geq 1$ (relatively higher dissipation) from the JHTD data ($4096 \times 4096 \times 192$ coarsened to $512\times 512\times 24$ tiles), superimposed with the $\Phi \leq 0.5$ field, which being spatially-exclusive, gives a clear impression that the most \textit{monofractal} flow regions are coincident with mild dissipation.}
	\label{fig:Local}
    \end{figure*} 

The dissipation field of course varies within these tiles, and we find it useful to keep track, for each tile, of the maximum $\epsilon_{\rm max}({\mathbf{x}})$, the minimum $\epsilon_{\rm
min}({\mathbf{x}})$ and mean  $\overline{\epsilon}({\mathbf{x}})$ dissipation as well as use 
$\Delta \epsilon ({\bf x}) \equiv \epsilon_{\rm max}({\mathbf{x}}) - \epsilon_{\rm min}({\mathbf{x}})$ as a measure 
of the fluctuation of the field; while all these values are presented as multiples of the global mean dissipation $\langle \epsilon \rangle$.
We are now equipped to calculate \textit{local} measures of multifractality ---
$D_q({\mathbf{x}})$ and $f_\alpha({\mathbf{x}}) - \alpha({\mathbf{x}})$ --- and estimate conclusively how uniformly 
(or not) multifractal turbulence really is. We construct a precise estimate of this through 
$\Phi({\mathbf{x}}) \equiv {\rm std}(\alpha({\mathbf{x}})) = \sqrt{ \langle \alpha^2 \rangle - \langle\alpha \rangle^2 }$, where $\langle .. \rangle$ denotes an average over all values of $\alpha$.  This provides a quantitative measure, using the spread of singularity strengths, of the degree of local multifractality in the flow. We know that these multifractal measures, even locally, should satisfy bounds such as 
$\alpha \geqslant \alpha_{\mathrm{min}}=-2$,
$f_{\alpha}\leqslant \alpha +2$, and $f_{\alpha }\geqslant 0$~\cite{dubrulle2022correspondence}.
If the $f_\alpha$ spectrum has a peak $f_\alpha^\star$ corresponding to some $\alpha = \alpha^\star$, then we have $f_\alpha^{\star} \lesssim 3$ (where $\alpha^{\star}=1$ and $f_\alpha^{\star}=3$ in the Kolmogorov framework). While there is no bound for $\alpha _{\mathrm{max}}$, it is reasonable to assume $\alpha _{\mathrm{max}} \approx 3$ for a region with no singular structures. Such monofractal regions can be expected to show $\Phi \approx 0$. However, the largest values of $\Phi$, 
corresponding to highly multifractal regions, is estimated as 
$\Phi \approx \sqrt{ \langle\alpha^2\rangle - \langle \alpha \rangle^2} \approx 1.7 \sim \mathcal{O}(1)$ 
for $\alpha$ uniformly ranging from $-2$ to $3$. Hence, on such theoretical grounds we expect 
$0 \lesssim \Phi({\bf x}) \lesssim 1.7$, with the lower and upper bound corresponding to mono and multifractal 
statistics, respectively.


In Fig.~\ref{fig:Local}(A) we show  $D_q ({\bf x})$ vs $q$ curves measured at different
spatial positions, corresponding to different values of $\epsilon_{\rm max}$. Clearly, while the shape of each curve is similar to
the global statistics (Fig.~\ref{fig:eps}, Inset A), a very strong spatial
dependence on where we measure the generalized dimensions is unmissable.
Furthermore, the spread in $D_q ({\bf x})$ is not trivially related to the
magnitude of maximum dissipation around ${\bf x}$; the secret to this variation, as we shall demonstrate, lies in how locally fluctuating (within each tile) the dissipation field is.

The measurement of the generalized dimension $D_q ({\bf x})$ allows us now to calculate local singularity spectra. In Fig.~\ref{fig:Local}(B), we show
representative plots of $f_\alpha ({\bf x}) - \alpha ({\bf x})$ for the same
locations (see legend in panel A) for which the generalized dimensions were
calculated. Quite clearly --- and contrary to what one sees in the conventional
global measurements of the singularity spectrum [see, e.g.,
Refs~\cite{Frisch-Book,MS-Nucl} and Fig.~\ref{fig:eps}, inset] --- there are
several regions where the flow is essentially monofractal (the $f_\alpha$ spectrum being very narrow) and fully consistent
with the ideas of Kolmogorov, while other highly multifractal regions lead to broad $f_\alpha$ curves. Furthermore, we find a very slight, but detectable drift to higher 
values of $\alpha$ where the $f_\alpha$ peaks as $\epsilon_{\rm max}$ increases. These results already hint that multifractality can be considered as a \textit{local} property of the field. 

In Fig.~\ref{fig:Local}(C) we show a pseudo-color plot of $\Phi$. Quite remarkably,
much of the flow is Kolmogorov-like with $\Phi \ll 1$; the highly
multifractal regions --- $\Phi ({\bf x}) \sim \mathcal{O}(1)$ --- are isolated patches
which, as we shall see, correlate completely with the extreme (singular) regions of energy
dissipation. This result is remarkable. It illustrates that, surprisingly, 
turbulent flows are \textit{not} uniformly multifractal; indeed on the contrary, much of the 
turbulent flow seems to respect, locally, Kolmogorov's ideas of an exact, self-similar cascade.
We also note that the range of $\Phi({\bf x})$ is well bounded by the theoretical range that we have 
discussed above.

What determines the magnitude and variation of $\Phi ({\bf x})$? Measurements
of the generalized dimensions and singularity spectra suggest that the
strength of the local dissipation  $\epsilon_{\rm max}$ is not where the answer
lies. We find that the probability distribution function
(pdf) of $\Phi$, conditioned on $\Delta \epsilon ({\bf x})$, is revealing.
In the inset of Fig.~\ref{fig:Local} (D) we show this pdf for three different values of $\Delta \epsilon$. Clearly, as
evident from the previous measurements, the distribution is sharply peaked at
values of $\Phi \gtrsim 0$ with an (likely) exponential tail for $\Phi \sim
\mathcal{O}(1)$. We also find that the probability of having a higher degree of
multifractality increases, albeit marginally, when there is a greater variation
of $\epsilon ({\bf x})$ within a tile. The mean value 
$\overline{\Phi}$, for a given $\Delta \epsilon$ (sampled in windows of $\Delta \epsilon \pm 0.25$), in fact grows logarithmically, as seen clearly in Fig.~\ref{fig:Local}(D). 

What then is the role of the average dissipation $\overline{\epsilon} ({\bf x})$ in determining the spatial non-uniformity of 
multifractality $\Phi ({\bf x})$? A joint-distribution (Fig.~\ref{fig:Local} (E)) shows that the answer is fairly non-trivial. Clearly, for low values of $\overline{\epsilon} ({\bf x})$ 
it is far more likely to have $\Phi ({\bf x}) \ll 1$; although, surprisingly, the less-likely extreme values of $\Phi ({\bf x})$ also coincide with regions of low $\overline{\epsilon} ({\bf x})$. This reflects that it is not the mean disspation in a region, but the \textit{variation} of dissipation, that manifests multifractality (as shown in Fig.~\ref{fig:Local} (D)). At higher $\overline{\epsilon} ({\bf x})$, the smallest values of $\Phi ({\bf x})$ admissible slightly increases with $\overline{\epsilon} ({\bf x})$, while the largest values of $\Phi({\bf x})$ also dip. While this result might appear contrary to our notion that extreme dissipation \textit{alone} begets multifractality, it finds parallel in an equally intriguing finding from a recent study showing local H\"older exponents, measured by proxy, also do not trivially correlate with inertial dissipation~\citep{dubrulle2019}. In fact, experiments have shown that the \textit{most} dissipative structures locally resemble Burgers vortices~\citep{debue2021three}. While these intense spots make the entire field highly intermittent and contribute to broadening the global multifractal spectrum, the \textit{local} multifractal picture can be different.

We finally cement these results with a visual illustration of where the Kolmogorov-like regions are embedded. We look at a snapshot (with volume-rendering) in Fig.~\ref{fig:Local} (F) of the 3D dissipation field, restricted to large values of $\overline{\epsilon}({\bf
x}) \ge 1$. Superimposed on this is the local measure of $\Phi ({\bf x}) \le 0.5$.
Unlike the sparsely populated high
$\overline{\epsilon}({\bf x})$ regions, the more frequent low $\overline{\epsilon}({\bf x})$ regions (hidden from view here) remain largely occupied by low $\Phi({\bf x})$ (these regions are also coincident with mild to low kinetic energy). Clearly, then, the regions of monofractal flow are \textit{strongly correlated} to the more populous regions of mild dissipation, showing that the Kolmogorov-like regions locally dissipate less than the multifractal regions.

\begin{figure}
	\includegraphics[width=1.0\columnwidth]{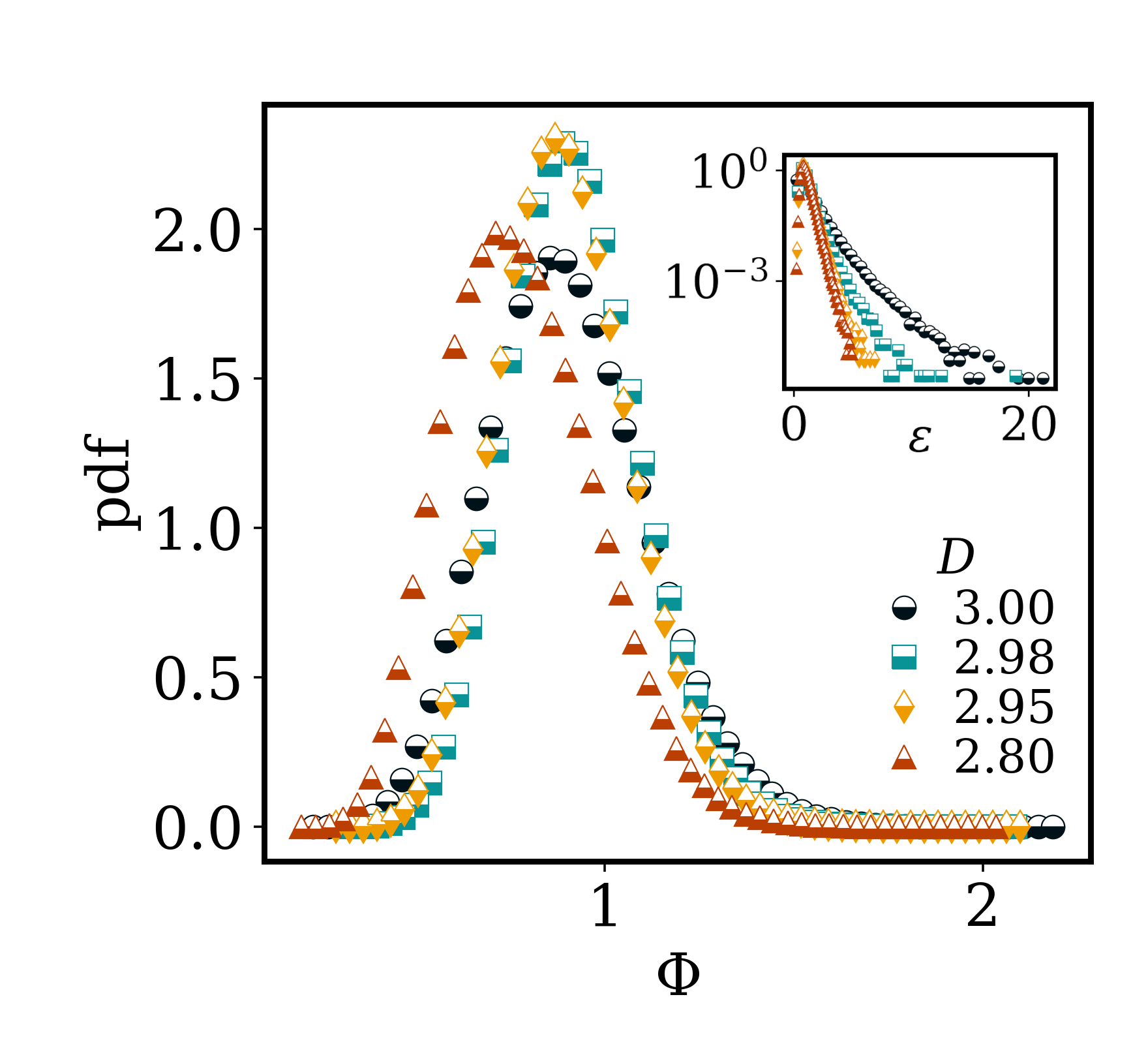}
	\caption{\textbf{Non-intermittent Turbulence.}
		Local multifractal statistics from decimated flow simulations. The inset shows that increasing decimation leds to a significant narrowing of the tails of the $\overline{\epsilon}$ pdfs. This is accompanied by the pdf of $\Phi$ itself shifting to lower values, reflecting reduced local multifractality corresponding to a reduction in intermittency.}
	\label{fig:decimated}
    \end{figure} 
    
We have, so far, shown compelling evidence which suggests that multifractality in turbulent flow
is not as spatially uniform as one might have suspected. In the absence of a
robust theory to explain this singular feature of turbulence, we make a final
test of these ideas in a Navier-Stokes-like flow which is \textit{guaranteed}
to be non-intermittent. An obvious choice for this is the so-called decimated
turbulence model which was introduced by Frisch \textit{et al.}~\citep{Frisch-PRL2012}. The basic principle lies in (numerically) solving
the Navier-Stokes equation on a Fourier lattice with a quenched disorder ---
namely the absence of a pre-chosen set of modes either randomly or fractally
--- by ensuring both the initial conditions and the non-linearity are projected
on this sub-set of remaining Fourier modes. Subsequent to the introduction of this model, 
we now know~\cite{Biferale-PRL,Ray-Review,Buzzicotti-NJP,Buzzicotti-PRE,Ray-PRF,Picardo-PRF} 
that such surgical removal of modes lead to a \textit{turbulence} 
which is non-intermittent.

We take advantage of such a flow to repeat the local multifractal analysis
performed on regular turbulence. A confirmation of the conclusions drawn from
our results would mean that the decimated flow ought to show lower values of $\Phi ({\bf x})$ than what is measured in fully developed
turbulence. Indeed, this is what we find in Fig.~\ref{fig:decimated}, showing
measurements of the pdf of $\Phi ({\bf x})$ for several different levels of
fractal decimation (which leads to decreasing intermittency as seen from the
pdfs of $\overline{\epsilon}$ in the inset). The $\Phi$ distributions
consistently shift toward lower values. Joint-distributions of $\Phi$ and
$\overline{\epsilon}$ were also found to show a simultaneous reduction in their spreads.
This confirms a strong link between intermittency and local multifractality. 

In conclusion, we wish to highlight two equally important contributions of this
paper. First is the finding that turbulence fields are not \textit{uniformly}
multifractal, but instead manifest strong multifractality in localized pockets
of intermittency in a quiescent Kolmogorovean background of mild dissipation. This paints over a
lacuna in our understanding of turbulence where, owing to the very construction
of the classical multifractal analysis, the notion of a spatially varying
multifractality remained inconcievable. Secondly, our local analysis framework opens
up a completely novel avenue for studying both the structure and dynamics of flow singularities and generalized dimension fields, in tandem with turbulence structures like intense vorticity worms~\cite{she1990intermittent,jimenez1993structure,moisy2004geometry}, non-locally induced velocity jets~\cite{mukherjee2022correlation}, or precursors to singular
dissipation~\cite{DubrullePRE,debue2021three}. This projects multifractality out of its role of
simply being a statistically reductive tool unrevealing of spatio-temporal minutiae, as noted before ~\citep{saw2016experimental}, to possible applications in prediction
and diagnostics of flows. Moreover, this localized analysis begs to
be applied to data from across disciplines, where multifractality has been
found emergent including in the areas of physics and
chemistry~\cite{stanley1988multifractal}, medicine~\citep{lopes2009fractal},
geophysics~\cite{mandelbrot1989multifractal}, climate~\citep{lovejoy2018weather} and
finance~\cite{jiang2019multifractal}, and is likely to be revealing, as we demonstrate for the long standing picture of turbulence, in unpredictable ways.

\begin{acknowledgements}
We are grateful to Rajarshi Chattopadhyay for help with obtaining the JHTDB data at \href{http://turbulence.pha.jhu.edu}{http://turbulence.pha.jhu.edu}. We thank Jason Picardo and J\'er\'emie Bec for insightful discussions and suggestions.
	S.S.R. acknowledges SERB-DST (India) projects MTR/2019/001553,
	STR/2021/000023 and CRG/2021/002766 for financial support. The authors 
	acknowledge the support of the DAE, Government of India,
	under projects nos. 12-R\&D-TFR-5.10-1100 and RTI4001.

\end{acknowledgements}

\bibliography{references}
\end{document}